\documentclass[a4paper,11pt]{article}
\usepackage{pos}
\usepackage{tikz}
\usepackage{float}

\definecolor{myblue}{HTML}{76D6FF}
\definecolor{bubblegum}{HTML}{FF85FF}

\title{Next-Generation Triggering:\\ A Novel Event-Level Approach}

\author*[a]{Jelena Köhler}
\author[f]{Aurélien Benoit-Lévy}
\author[b]{Pablo Correa}
\author[b,f]{Arsène Ferriere}
\author[a,d]{Tim Huege}
\author[c,e]{Kumiko Kotera}
\author[b]{Olivier Martineau-Huynh}
\author[c,g]{Simon Prunet}
\author[a]{Markus Roth}

\onbehalf{for the GRAND collaboration}

\affiliation[a]{Karlsruhe Institute of Technology (KIT), Institute for Astroparticle Physics, Karlsruhe, Germany}

\affiliation[b]{Sorbonne Université, Université Paris Diderot, Sorbonne Paris Cité, CNRS, Laboratoire de physique nucléaire et des hautes énergies (LPNHE), 4 place Jussieu, F-75252, Paris Cedex 5, France}
\affiliation[c]{Sorbonne Université et CNRS, UMR 7095, Institut d’Astrophysique de Paris, 98 bis bd Arago, 75014 Paris, France}
\affiliation[d]{Vrije Universiteit Brussel (VUB), Astrophysical Institute, Brussels, Belgium}
\affiliation[e]{Vrije Universiteit Brussel (VUB), Dienst ELEM, Pleinlaan 2, B-1050, Brussels, Belgium}
\affiliation[f]{Université Paris-Saclay, CEA, List, F-91120, Palaiseau, France}
\affiliation[g]{Université Côte d’Azur, Observatoire de la Côte d’Azur, CNRS, Laboratoire Lagrange, Bd de l’Observatoire, CS 34229, 06304 Nice cedex 4, France}

\emailAdd{jelena.koehler@kit.edu}

\abstract{
Large-scale cosmic-ray detectors like the Giant Radio Array for Neutrino Detection (GRAND) are pushing the boundaries of our ability to identify air shower events. 
Existing trigger schemes rely solely on the timing of signals detected by individual antennas, which brings many challenges in distinguishing true air shower signals from background. 
This work explores novel event-level radio trigger methods specifically designed for GRAND, but also applicable to other systems, such as the Radio Detector (RD) of the Pierre Auger Observatory. 
In addition to an upgraded plane wave front reconstruction technique, we introduce orthogonal and complementary approaches that analyze the radio-emission footprint, the spatial distribution of signal strength across triggered antennas, to refine event selection. 
We test our methods on mock data sets constructed with simulated showers and real background noise measured with the GRAND prototype, to assess the performance potential in terms of sensitivity and background rejection in GRAND. 
Our preliminary results are a first step to identifying the most  discriminating radio signal features at event-level, and optimizing the techniques for future implementation on experimental data.
}

\FullConference{%
  10th International Workshop on Acoustic and Radio EeV Neutrino Detection Activities - ARENA2024\\
  10-14 June 2024\\
  University of Chicago, Chicago, USA
}

\begin{document}
\maketitle

\section{Introduction: Towards Autonomous Triggering with GRAND}
The Giant Radio Array for Neutrino Detection (GRAND) is an emerging observatory designed to detect ultra-high energy neutrinos and cosmic rays with unparalleled sensitivity. 
It consists of a sparse radio antenna array that covers an area on the order of hundreds of kilometers, allowing it to identify particle showers in the Earth's atmosphere. 
The deployment of the first prototype, known as GRANDProto300 (GP300), was initiated in the spring of 2023 near Dunhuang, China. 
The initial installation consisted of 13 antennas, which will be steadily expanded to a total of 300 antennas in the near future.
A key innovation of GRAND is its autonomous triggering system. 
Unlike previous radio experiments that decided to rely on external triggers, GRAND's antennas will independently detect and record radio signals from air showers. 
This capability is essential to efficiently monitor the vast areas required to capture high-energy cosmic ray and neutrino events.

In the scope of the NUTRIG project, a multi-level trigger system is being developed for GRAND, with the goals of high purity, high efficiency and good scalability. 
A high-purity trigger system minimizes false positives, so that only genuine air showers are flagged for further analysis. 
A high efficiency takes this a step further and reduces the number of false negatives (i.e. missed true events).
Scalability ensures that the trigger system will be able to handle increasing amounts of data while maintaining performance and reliability.
It is essential for accommodating the expanding size of the GRAND array as well as confirming the adaptability to other experiments.
For each antenna, an FPGA-based first-level trigger (FLT-0) is followed by a CPU-based counterpart (FLT-1). 
Together, they are simply referred to as FLT.
This work focuses on developing the second-level trigger (SLT), which acts on event level. 
Offline analysis has shown that crucial shower characteristics can be condensed into a few key parameters: trigger time~\cite{arsene}, signal strength and polarization~\cite{Chiche:polarization}. 
The goal of this research is to create efficient algorithms for implementation in a real-time ``online'' trigger.
To optimize data handling, the proposed trigger system aims for specific rate reductions. 
While GP300 will nominally operate with a $1\,$kHz FLT and $10\,$Hz SLT trigger rate, the NUTRIG project targets a more stringent $100\,$Hz for the FLT and 1 Hz for the SLT. 
These rate reductions are crucial for efficient data management and analysis. 
We present the methodology for both timing-based and signal strength-based event selection techniques in Section~\ref{sec:method}. The performance of these methods is evaluated using a realistic mock dataset, as described in Section~\ref{sec:application}. Finally, a summary of the results and an outlook on future work are provided in Section~\ref{sec:summary}.

\section{Methodology}\label{sec:method}

To effectively identify air shower events, it is crucial to establish their distinct characteristics. 
This research focuses on two orthogonal methods: a timing analysis for accurately determining the event's arrival direction~\cite{arsene}, and a signal strength analysis to recognize the characteristic radio footprint geometry of the air shower on the antenna array.

\subsection{Timing: Plane Wave Front Reconstruction}

To estimate the arrival direction of cosmic ray showers, we employ a plane wave front (PWF) reconstruction method. 
This technique, commonly used in radio detection experiments, approximates the shower front as plane and determines its arrival direction based on the arrival time of signal traces from individual antennas. 
We employ a novel implementation of the PWF method \cite{arsene}, that provides analytical solutions for the shower arrival direction and its uncertainty. 
Because it is fast (no iterative fitting necessary) and robust, this reconstruction method can be used as a proxy for more complex estimators, or implemented online for low-level triggering. 
Here we test the method for the first time on mock datasets constructed from simulated voltage traces complemented with measured noise.

\subsection{Signal Strength: Elliptical Footprint Geometry Recognition}

\begin{figure}[H]
    \centering
    \includegraphics[width=0.6\linewidth]{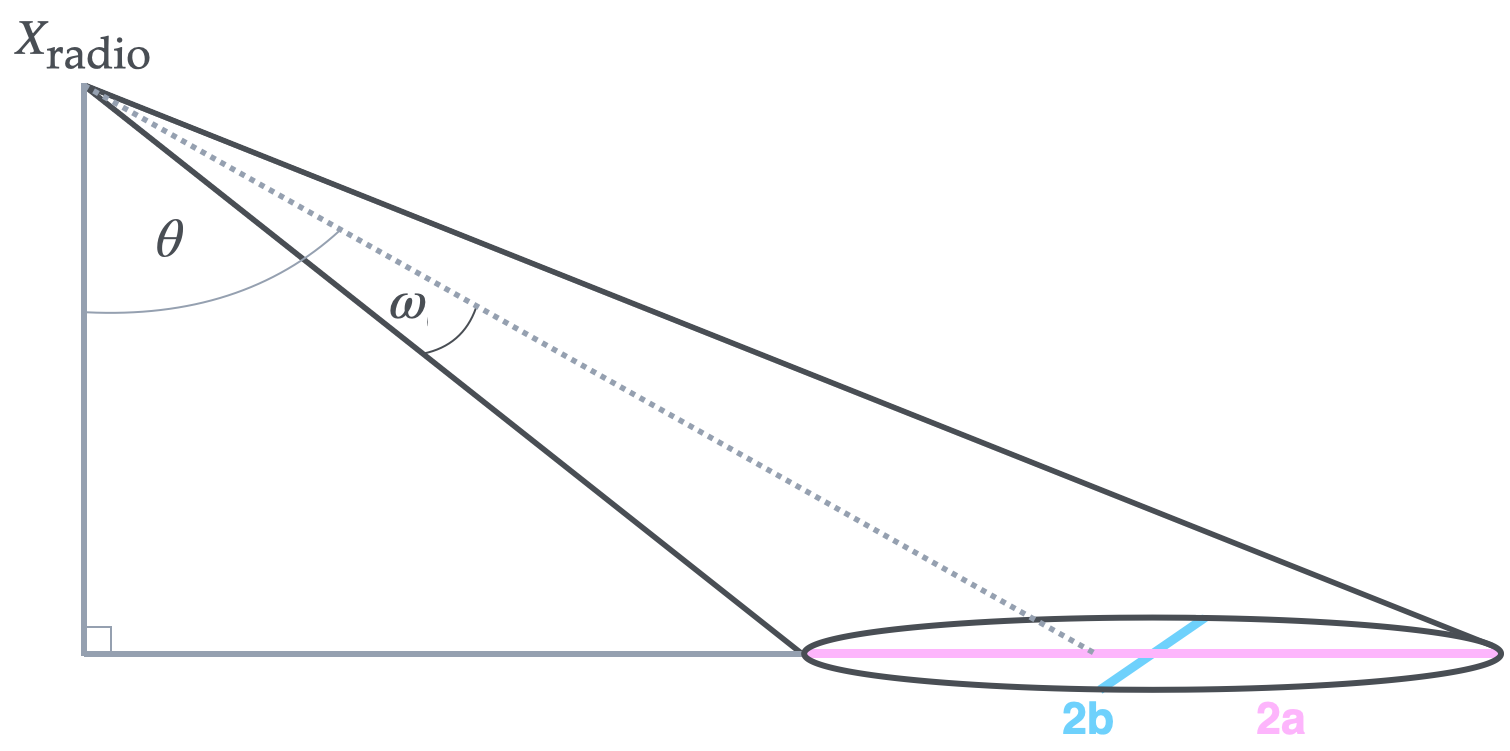}
    \caption{A sketch of an inclined air shower with labeled geometrical properties. The dotted line indicates the shower axis. The radio signal is modeled as a cone with opening $\omega$, leading to an elliptical footprint on the ground.}
    \label{fig:triangle}
\end{figure}

A suitable model for the footprint of an air shower involves the assumption that it corresponds to the intersection of a cone (the radio signal beam) of opening angle $\omega$ with a plane (the ground), resulting in an ellipse (cf.~Fig.~\ref{fig:triangle}). 
The radio signal beam is assumed to be mostly emitted from a point $X_{\rm radio}$ that corresponds roughly to the atmospheric column density $X_{\rm max}$ at which the shower reaches its maximum particle number. 
The initial geometrical properties of the shower (zenith angle $\theta$ and azimuth angle $\varphi$) can be related to the ellipse parameters $a$ and $b$, the semi-major and semi-minor axes respectively. 
In particular, the eccentricity $\varepsilon$ of the ellipse allows us to quantify the spatial characteristics of the shower footprint and put it in context with the zenith angle $\theta$ and azimuth angle $\varphi$ of the shower.
The azimuth angle is directly related to the orientation of the ellipse.
A low eccentricity corresponds to a more circular shape and a small zenith angle $\theta$, while a high eccentricity indicates a more elongated form and a large zenith angle $\theta$. 
It is defined as $\varepsilon = {\sin\theta}/{\cos\omega} = {\sqrt{a^2-b^2}}/{a}$.
The opening angle $\omega$ can be approximated as three Cherenkov radii corresponding to $3\omega_{\rm C}\sim 1.5^\circ$ for inclined air showers, cf.\ reference \cite{PierreAuger:2018pmw}. The zenith angle can thus be calculated as
\begin{equation}\label{eq:zenith}
    \sin\theta = \frac{\sqrt{a^2-b^2}}{a}\cdot\cos(3\omega_{\rm C})
\end{equation}
This formula\footnote{We cross-checked this calculation with a simple $\cos\theta$ projection from the shower plane to the ground plane, yielding agreement within $0.01\%$.} provides a convenient way to determine the zenith angle of an air shower event based on the geometric properties of an ellipse that can be fitted to the radio-emission footprint.

In order to analyze the radio-emission footprint, we introduce the ``ADC fluence''. 
It is calculated as the sum of the time integrals of the squared voltages for the east-west and north-south antenna channels for the full trace length of $\approx8\,000\,\text{ns}$.
It provides a rapid approximation of the energy fluence, albeit convolved with the antenna response. 
Due to the negligible contribution of the vertical component, it is excluded for computational efficiency.
We perform a principal component analysis on the antenna positions weighted by their ADC fluences to determine the orientation and size of a footprint \cite{PCA}.
First, the antenna $x$ and $y$ positions are combined into a single matrix, and weights based on their ADC fluence are calculated. 
A weighted covariance matrix is computed to capture the distribution of antennas considering their relative importance. 
Eigenvalue decomposition of this covariance matrix yields the eigenvectors of the principal components, representing the directions of maximum variance which we identify with the two semi-axes of the ellipse.
The eigenvalues correspond to the variances along these components.
Specifically, we take the square root of the eigenvalues as an estimate of the length of the semi-axes, which is then used to calculate the shower zenith angle $\theta$ using Eq.~\ref{eq:zenith}.
Finally, the orientation of the footprint and thus the azimuth angle $\varphi$ can be directly obtained from the eigenvectors as $\varphi = \arctan(dy/dx)$, where $dy$ and $dx$ represent the components of the dominant eigenvector along the $y$ and $x$ axes, respectively.

\section{Application on Realistic Mock Data}\label{sec:application}
To evaluate the performance of the proposed methods under realistic conditions, we apply them to a realistic data set of simulated air showers. This section describes the simulation library used in this work and presents the results obtained from it for both the PWF and signal strength reconstruction techniques.

\subsection{Simulation Library}
For this analysis, we use a database of simulated air shower pulses for the GP300 array, created using ZHaireS~\cite{zhaires}, with varying energies, zenith angles, and azimuth angles. 
The dataset contains about $7\,000$ air shower events with energies $\log_{10}(E/\rm{eV})\in [16.5,18.5]$, zenith angles $\theta\in [30.6^\circ,87.3^\circ]$ sampled in $\cos\theta$, and azimuth angles $\varphi\in [0^\circ,360^\circ]$ sampled uniformly and randomly. 
This dataset is based on the simulation library built in the framework of GRAND reconstruction efforts, which incorporates $200\,000$ air showers~\cite{oscar}. 
One can note that the mock GP300 array utilized in this study predates the official determination of GP300 antenna positions~\cite{simonARENA}, but suffices for assessing the performance of the method. 
The spacing between antennas and the unitary hexagonal pattern are identical, the differences are mainly in the global shape of layout and the denser infill.

In order to avoid too small footprints, we filter for a minimum number of five antennas to ensure sufficient spatial resolution. 
On the other hand, highly inclined air showers will become too large for the simulated GP300 array and border effects will invalidate the analysis.
Furthermore, the bulk of measured statistics will not be at such high zenith angles in real data. 
Therefore, we decide to only use zenith angles up to $82.5^\circ$ in this analysis.

The simulations are processed with the GRANDlib software~\cite{grandlib} to replicate the complete radio-frequency chain. 
For this work, our specific library further includes measured background noise from the currently deployed thirteen GP300 antennas under first-level trigger conditions (cf.~\cite{pablo}). 
This approach creates a realistic representation of air shower signals as observed by the array, although heavily weighted towards inclined showers.
We note that neither timing uncertainties (mimicking the limited accuracy of GPS clocks) nor antenna-to-antenna signal strength fluctuations (mimicking unit-to-unit variations) have been imprinted at this stage of the analysis.

\begin{figure}
     \centering
         \centering
         \includegraphics[width=0.48\linewidth]{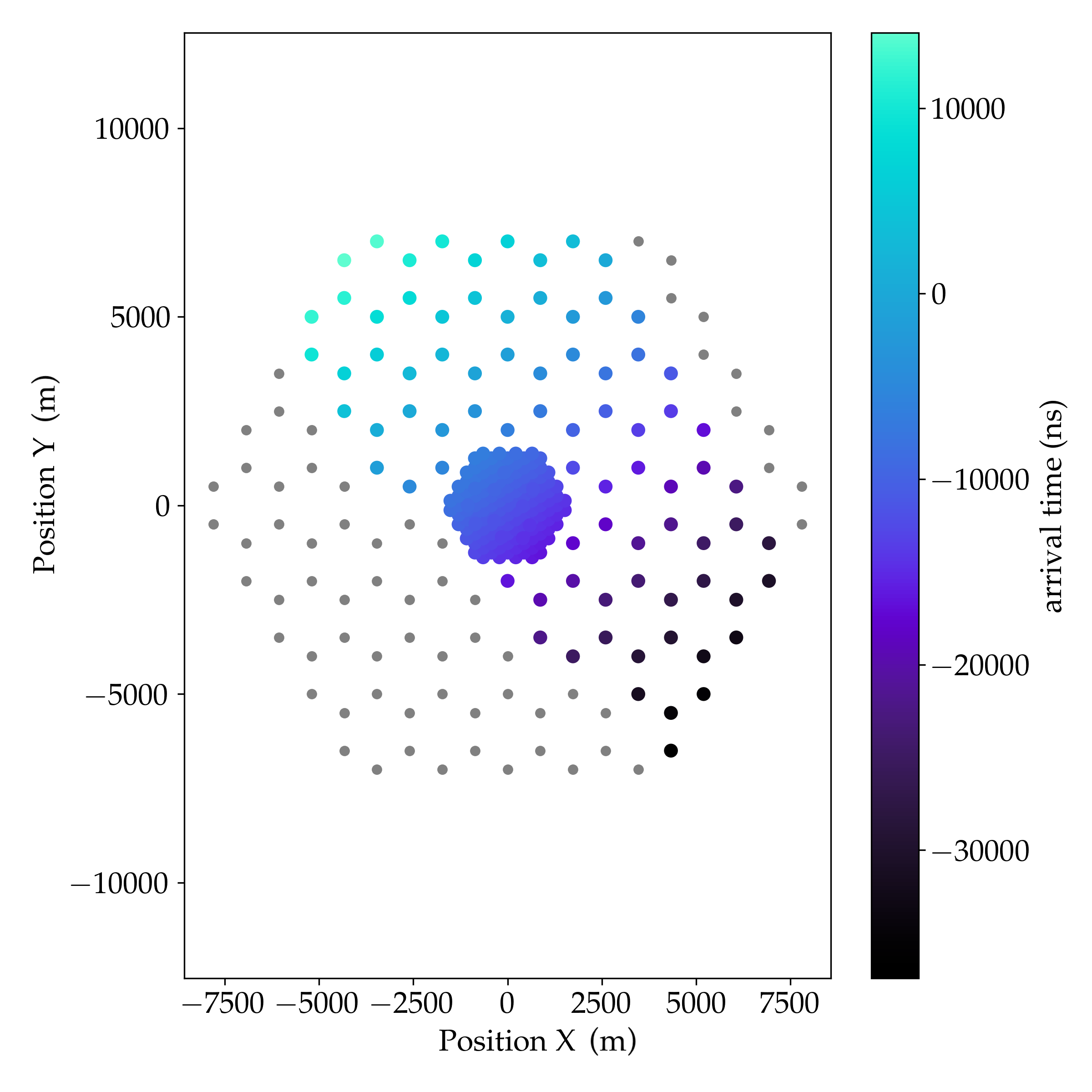}
        \includegraphics[width=0.48\linewidth]{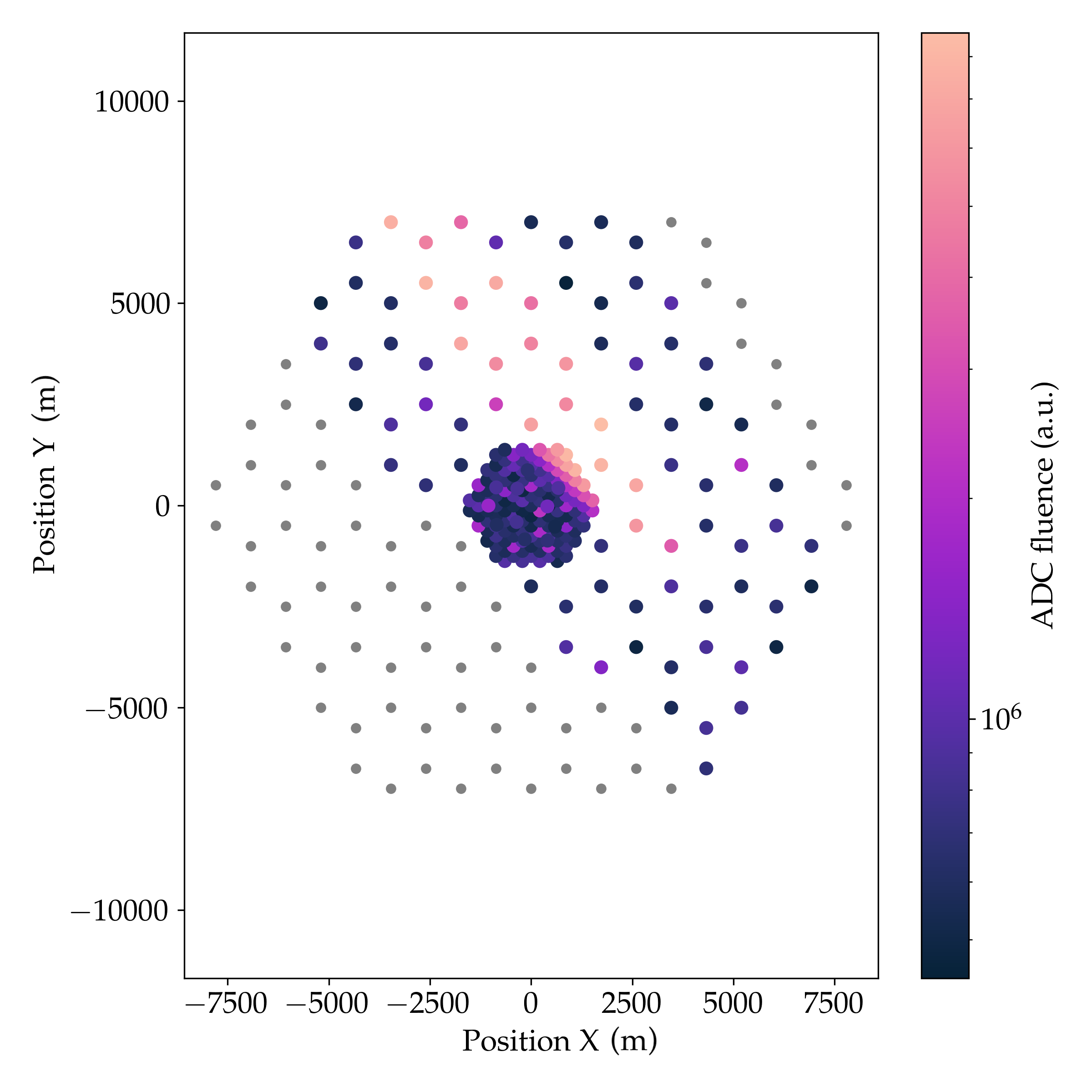}
        \caption{ 
        Arrival times in nanoseconds {\it (left)} and ADC fluence in arbitrary units {\it (right)} at the antenna positions of a mock GP300 layout, for a simulated shower event of primary proton of energy $E = 10^{18.9}\,$eV, zenith angle $\theta = 80.7^\circ$, azimuth angle $\varphi = 308.5^\circ$. The ellipse of the radio-emission footprint is noticably cut for this geometry.}
        \label{fig:GP300event}
\end{figure}

An example event is shown in Fig.~\ref{fig:GP300event}, where the two plots (left and right) present the arrival times and ADC fluence distributions across the GP300 array, respectively. 
The arrival time  shows that in this particular example, the shower originated from the bottom right $x$-$y$ quadrant and propagated towards the positive $x$-$y$ top left.
The expected radiation pattern is clearly visible in the ADC fluence map, providing a distinctive footprint of the event.
For more information on the simulation library, refer to~\cite{pablo}.

\subsection{Plane Wave Reconstruction - Results}
In order to apply the plane wave reconstruction from \cite{arsene}, we retrieve the antenna positions  and the time stamps from the simulation library.
The time stamps represent the FLT trigger time, which are given in GPS seconds and GPS nanoseconds.

\begin{figure}[H]
     \centering
        \centering
        \includegraphics[width=0.49\linewidth]{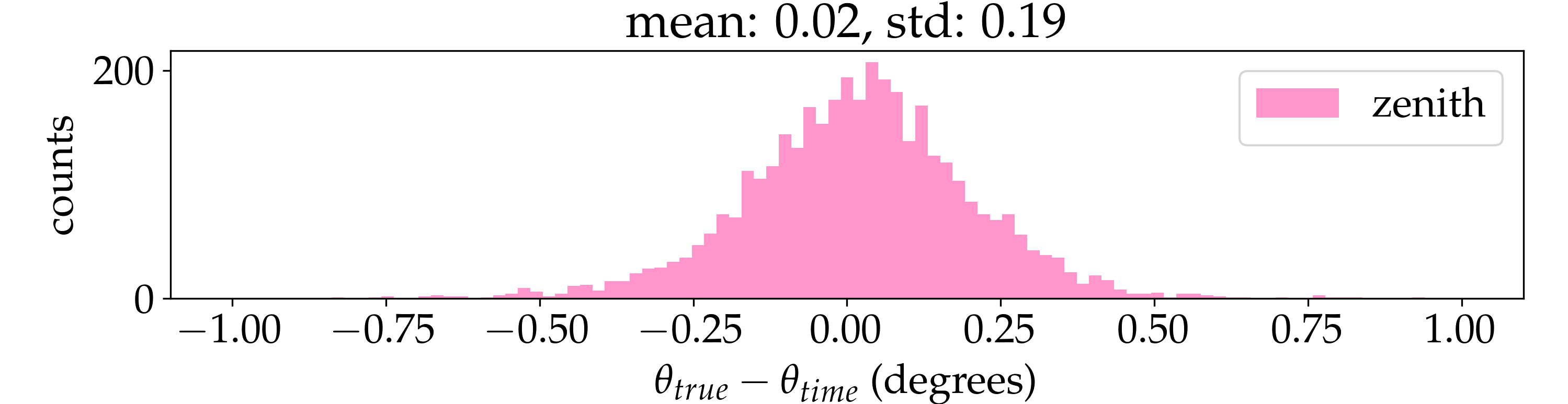}
        \includegraphics[width=0.49\linewidth]{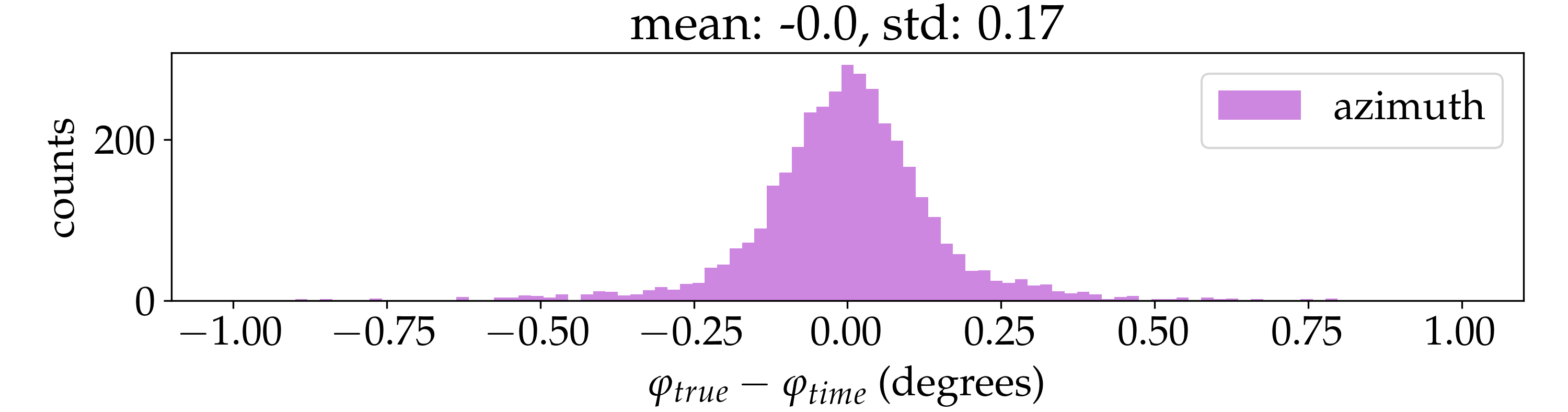}
        \caption{Histograms of the difference between the Monte Carlo true and the reconstructed angles using the plane wave reconstruction method.}
\label{fig:arrivaltimes_comparison_hist}
\end{figure}

The results of the PWF reconstruction are shown in
Fig.~\ref{fig:arrivaltimes_comparison_hist}. 
The left histogram shows the difference between the Monte Carlo truth of the zenith angle $\theta_{\rm true}$ and the reconstructed zenith angle $\theta_{\rm time}$ in degrees. 
The right histogram shows the difference between the Monte Carlo truth of the azimuth angle $\varphi_{\rm true}$ and the reconstructed azimuth angle $\varphi_{\rm time}$ in degrees. 
For both we get symmetrical Gaussian distributions centered around zero, with means~$\approx 0.0^\circ$ and standard deviations below~$0.2^\circ$. 
While this analysis provides valuable insights, these unrealistically low deviations between true and reconstructed angles suggest that they may not adequately represent the challenges and uncertainties that are to be expected in experimental data.

\subsection{Signal Strength Reconstruction - Results}
In order to analyze the signal strength distribution across the array, we retrieve the antenna positions and the ADC fluence of each antenna from the simulation library and perform the above-described principal componant analysis. 

\begin{figure}[H]
     \centering
        \centering
        \includegraphics[width=0.49\linewidth]{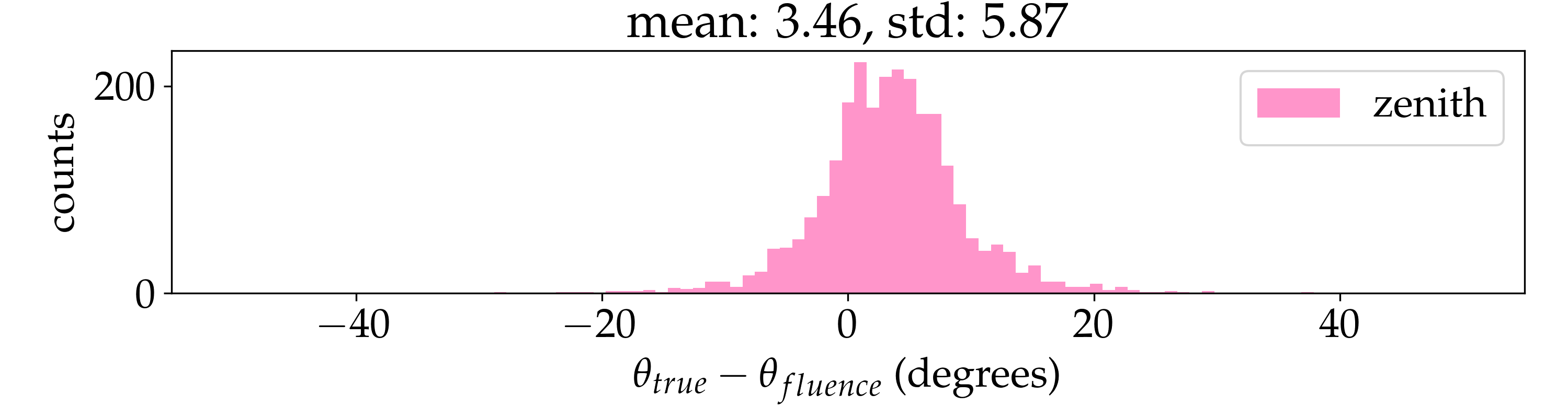}
        \includegraphics[width=0.49\linewidth]{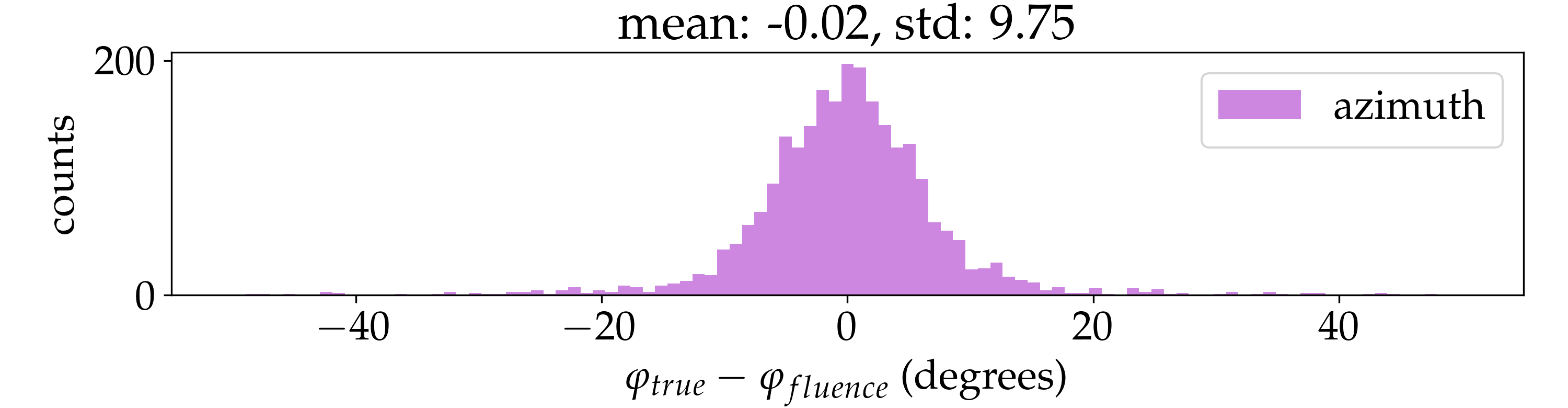}
        \caption{Histograms of the difference between the Monte Carlo true and the reconstructed angles using the signal strength method.}
\label{fig:signalstrengths_comparison_hist}
\end{figure}

\begin{figure}[H]
    \centering
   \includegraphics[width=0.6\linewidth]{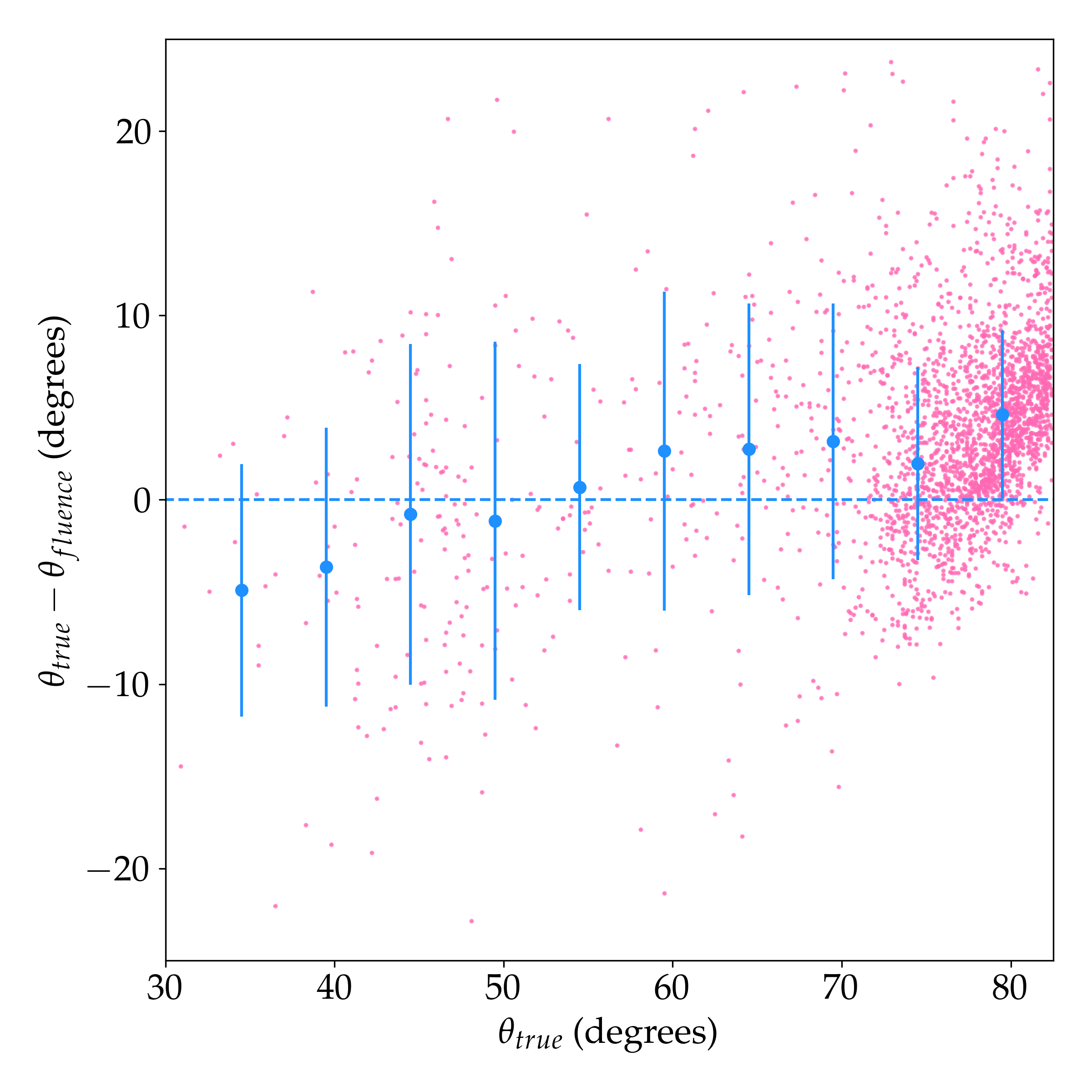}
    \caption{Scatter plot comparing the true zenith angles to the difference between the true and reconstructed zenith angles. 
    Each pink point represents one event, and the blue dashed line marks the expected ideal distribution. 
    The blue points describe the distribution's profile, including statistical uncertainties in terms of the standard deviation.}
\label{fig:signalstrength_comparison_scatter}
\end{figure}

The results of this method are shown in
Fig.~\ref{fig:signalstrengths_comparison_hist}. 
The left histogram shows the difference between the Monte Carlo truth of the zenith angle $\theta_{\rm true}$ and the reconstructed zenith angle $\theta_{\rm fluence}$ in degrees. 
The right histogram shows the difference between the Monte Carlo truth of the azimuth angle $\varphi_{\rm true}$ and the reconstructed azimuth angle $\varphi_{\rm fluence}$ in degrees. 
Without using the timing information, a $180^\circ$ ambiguity on the determination of the azimuth angle arises.
Applying a $\rm{mod}\,180^\circ$ operation, we get a symmetrical Gaussian distribution centered around zero, with a mean of~$-0.02^\circ$ and a standard deviation of $9.75^\circ$, for the azimuth angle distribution. 
However, the zenith angle distribution shows a 
bias with a mean of $3.46^\circ$ and a standard deviation of $5.87^\circ$. 
In order to investigate this bias further, Fig.~\ref{fig:signalstrength_comparison_scatter} shows a scatter plot comparing the true zenith angles to the difference between the true and reconstructed zenith angles.
The dashed line represents one-to-one agreement, serving as a benchmark for comparison. 
We overlay the scatter plot with a profile showing the mean values and standard deviations in bins of $5^\circ$.
The plot indicates that the number of data points strongly increases towards higher zenith angles, starting from $\theta_{\rm true}\approx 70^\circ$. 
This biased statistic might influence the analysis. 

Several other factors could also explain the observed zenith angle bias. 
The mathematical approximations underlying the reconstruction method may break down at higher zenith angles, resulting in increased inaccuracies. 
Furthermore, edge effects occurring at the boundaries of the simulated antenna array are very likely to be a possible cause. 
Especially large footprints that are not fully visible on the detector array (cf.~Fig.~\ref{fig:GP300event}) can be difficult to evaluate with the current method. 
A potential mitigation strategy could be to successively apply increasing ADC fluence thresholds and check whether the reconstructed geometry converges to a stable result.

Nevertheless, it is important to emphasize that the primary objective of this method is to efficiently identify potential air shower events at trigger level rather than achieving precise reconstruction. 
A more comprehensive evaluation of the method performance will become possible once it is applied in addition to real-world self-triggered noise data, thus allowing it to quantify efficiency and purity of the trigger decision.

\section{Summary and Outlook}\label{sec:summary}
This study introduces a novel method for analyzing signal strength distributions to enhance event triggering capabilities of self-triggered radio detectors. 
While this study was performed for the GP300 array, the derived method demonstrates promising potential for event triggering for any large-scale radio array. 
By combining an approach based on the evaluation of signal strengths with an existing plane wave fit reconstruction, we effectively characterize radio-emission footprint properties and identify potential air shower events. 
While the signal-strength reconstruction of azimuth angles proved unbiased, a systematic bias was found in the zenith angle reconstruction.
In a next step, a comprehensive evaluation of the method performance will be made with a library of self-triggered noise events. 
Future work will focus on addressing the observed zenith angle bias.
Since the main cause seem to be footprints that are not fully visible on the detector, refining the reconstruction method towards handling these cases should improve the results.
Partial footprints could be approximated as ellipses by introducing an ADC fluence threshold in the fitting procedure.
Furthermore, investigating the size of the signal footprint and its relation to the zenith angle, the distribution of the signal strength along the orthogonal axes of the footprint, and the measured signal polarization in comparison with the expectation \cite{Chiche:polarization}, could provide valuable insights into shower characteristics and improve trigger efficiency and purity.

\section*{Acknowledgements}
This work is part of the NUTRIG project, supported by the Agence Nationale de la Recherche (ANR-21-CE31-0025; France) and the Deutsche Forschungsgemeinschaft (DFG; Projektnummer 490843803; Germany). 
Computations were performed using the resources of the CCIN2P3 Computing Center (Lyon/Villeurbanne, France), a partnership between CNRS/IN2P3 and CEA/DSM/Irfu, and the HoreKa supercomputer, funded by the Ministry of Science, Research and the Arts Baden-Württemberg and by
the Federal Ministry of Education and Research.

\bibliographystyle{JHEP}
\bibliography{lit.bib}

\end{document}